\documentclass[11pt,a4paper]{article}
\usepackage{jheppub}
\usepackage{amsfonts}
\usepackage{amssymb}

\newcommand{\bea}{\begin{eqnarray}}
\newcommand{\eea}{\end{eqnarray}}
\newcommand{\bean}{\begin{eqnarray*}}
\newcommand{\eean}{\end{eqnarray*}}
\newcommand{\nn}{\nonumber\\}
\newcommand{\Sl}{\sum\limits}
\def\W #1{\widetilde{#1}}
\def\WH #1{\widehat{#1}}
\allowdisplaybreaks
\title{A Proof of the Explicit Minimal-basis Expansion of  Tree Amplitudes in Gauge Field Theory}

\author[a]{Yi-Xin Chen}
\author[a,c]{Yi-Jian Du}
\author[b,d]{Bo Feng}
\affiliation[a]{Zhejiang Institute of Modern Physics, Zhejiang University,\\
Hangzhou, 310027, P. R. China} \affiliation[b]{Center of
Mathematical Science, Zhejiang
University,\\
Hangzhou, 310027, P. R. China}
\affiliation[c]{Kavli Institute for Theoretical Physics China,Chinese Academy of Sciences,\\
Beijing 100190, P. R. China}
 \affiliation[d]{Key Laboratory of
Frontiers in Theoretical Physics, Institute of Theoretical Physics,
Chinese Academy of Sciences,\\Beijing 100190, P. R. China}

\abstract{ In last couple years, an important relation (BCJ
relation) between color-ordered tree-level scattering amplitudes of
gauge theory has inspired many studies. This relation implies that
the minimal basis for the color-ordered tree-level amplitudes is
$(n-3)!$ and other amplitudes can be expanded into a particular
chosen basis. In this paper we will prove the conjectured explicit
minimal basis expansion. For this purpose
we will write down general  BCJ relation of gauge theory by taking
 the field theory limit of BCJ relation in string theory. Then we prove these
 general BCJ relations using BCFW on-shell recursion relation. Using
 these general BCJ relations, we prove the
 conjectured  explicit minimal-basis expansion of gauge theory tree amplitudes
 inductively.

} 
\begin{document}
\maketitle

\section{Introduction}
Recently, there are  significant progresses about various relations
of tree-level scattering amplitudes of gauge theory and gravitational
theory. For gauge theory, old result
(Kleiss-Kuijf(KK)\cite{Kleiss:1988ne} relation) stated that any
color-ordered tree-level amplitude of $n$ gluons can be expanded by
a basis with $(n-2)!$ amplitudes. However, this result has been
revised after Bern, Carrasco and Johansson conjectured a new highly
nontrivial relation(BCJ relation)\cite{Bern:2008qj} for gauge theory
which significantly reduces the number of basis from $(n-2)!$ to
$(n-3)!$\footnote{The BCJ relation for amplitudes with gluons coupled to matter is suggested in \cite{Sondergaard:2009za}.}.
The new discovered BCJ relation plays a very important
role for our understanding of another important old result in
gravity theory ( Kawai-Lewellen-Tye(KLT)\cite{Kawai:1985xq}
relation) which expresses  tree-level amplitude of  $n$ gravitons by
the sum of products of two color-ordered tree-level amplitudes of
$n$ gluons with appropriate kinematic
factors\cite{Berends:1988zp,Bern:1998ug}.

These relations have been investigated from the point of view of
string theory as well as  field theory. From the point of view of
string theory,
 KK relation, BCJ relation\cite{BjerrumBohr:2009rd,Stieberger:2009hq}
 and KLT relation\cite{Kawai:1985xq,BjerrumBohr:2010hn}
are consequences of  monodromy relations. After taking the field
theory limit in string theory, these relations appear naturally.

Although it is convenience to embed  gauge  theory and gravity
theory  into  string theory, it is not necessary to do so. In fact,
as the consistent consideration, it is desirable to have a pure
field theory proof of these facts. The old result (KK relation) has
been proved  by new color-decomposition \cite{DelDuca:1999rs}. The
new discovered  fundamental BCJ relation (as well as KK relation)
and KLT relation have been proved by
Britto-Cachazo-Feng-Witten(BCFW) on-shell recursion
relation\cite{Britto:2004ap,Britto:2005fq}
 along the line of S-matrix program\cite{S-matrix-program} in \cite{Feng:2010my,
Jia:2010nz} and
\cite{BjerrumBohr:2010ta,BjerrumBohr:2010zb,BjerrumBohr:2010yc,Feng:2010br}\footnote{In
\cite{Tye:2010kg} BCJ relation is explained from the point of view
of Schouten identity.}. From the point of view of these proofs, the
BCJ relation is the bonus relation of the improved vanishing
behavior for non-nearby BCFW-deformation \cite{ArkaniHamed:2008yf}.
BCJ relation is also the consistent condition for the equivalence of
various  KLT relations.

BCFW recursion relation\cite{Britto:2004ap,Britto:2005fq} is an
important tool to calculate and study  amplitudes.
 With BCFW recursion relation,
one can construct on-shell tree amplitudes by sub-amplitudes with
less external legs
\bea
M_n=\Sl_{\mathcal{I},\mathcal{J},\mathbf{h}}\frac{M_{\mathcal{I}}(\WH
p_i,\WH P_{\mathcal{I},\mathcal{J}}^{\mathbf{h}})M_{\mathcal{J}}(\WH
p_j,-\WH
P_{\mathcal{I},\mathcal{J}}^{-\mathbf{h}})}{P_{\mathcal{I}\mathcal{J}}},
\label{BCFW} \eea
where the sum is over all possible distributions of  external legs
with shifted momentum $\WH p_i\in\mathcal{I}$, $\WH
p_j\in\mathcal{J}$ and $z_{\mathcal{I}\mathcal{J}}$ indicates the
location of deformation where inner propagator is on-shell. Via
discussions on complex analysis, this relation exists in theories
with proper vanishing behavior $M(z\rightarrow\infty)=0$ under
BCFW-deformation. Both gauge  theory and gravity theory satisfy this
condition\cite{Britto:2005fq,ArkaniHamed:2008yf,z-infinite}. BCFW
recursion relation can also be written down with nontrivial boundary
contributions as considered in \cite{BCFW:boundary}\footnote{BCFW
recursion relation has been generalized to loop level as well as
string theory. A few references can be found, for example, in
\cite{BCFW:loop-level} and\cite{BCFW:string}.}.

Because its importance,  there have been lots of works on BCJ
relation. From our point of view, we think that
 BCJ relation can be understood from following two levels of meaning.
 The first level of meaning is
that there is a set of constraint equations which reduce the number
of independent amplitudes from $(n-2)!$ to $(n-3)!$.
 The second level of meaning is the explicit minimal-basis expansion of
 amplitudes, i.e., how other amplitudes can be written down as the linear
 combination of basis with explicit expression of coefficients.
For the first level of claim, in string theory, these constraints
come from monodromy relations when we do the contour deformation and
they contain not only the fundamental BCJ relation, but also other
relations which we will call ``general BCJ
relations''\cite{BjerrumBohr:2009rd,Stieberger:2009hq}. The
fundamental BCJ relation has been proved in field theory by BCFW
recursion relation, but the general BCJ relations have not been
proved and we will give a proof in this paper. For the second level
of claim, until now it is still a conjecture \cite{Bern:2008qj} and
there is  no explicit proof. It is our main purpose in this paper to
give such a proof, thus complete the whole claim.

The outline of our paper is following. In section \ref{F-T-limit} we
will first consider the field theory limit of the general BCJ
relation in string theory, which provides a set of constraints on
gauge theory amplitudes. It reduces the number of the independent
amplitudes from $(n-2)!$ to $(n-3)!$.
 Though the general  BCJ relation can be derived from string theory directly,
 we give also a field theory proof by BCFW recursion relation in section
\ref{second-proof}. These general BCJ relations will be used for the
proof of the explicit minimal-basis expansion of  gauge theory
amplitudes in section \ref{first-proof}. Finally in section
\ref{conclusion} we give a brief conclusion.

\section{The field theory limit of the BCJ relation in string
theory}\label{F-T-limit}

 Gauge field theory can be embed into open  string theory as its massless
 field limit. String theory often provides many useful information
for studying gauge theory. One of such examples  is the beautiful
proof of KK and BCJ
relations\cite{BjerrumBohr:2009rd,Stieberger:2009hq} in string
theory, where  monodromy plays a crucial role in the proof. In this
section, we will take the field theory limit of the BCJ relation in
string theory.

 KK relation for open string tree amplitudes  is given
 as\cite{BjerrumBohr:2009rd}
\bea & &A_n(\beta_1,...,\beta_r,1,\alpha_1,...,\alpha_s,n)\nn
&=&(-1)^r\times \mathcal{R}e\left[\prod\limits_{1\leq i<j\leq
r}e^{2i\pi\alpha'k_{\beta_i}\cdot k_{\beta_j}}\Sl_{\{\sigma\}\in
P(O\{\alpha\}\cup
O\{\beta^T\})}\prod\limits_{i=0}^s\prod\limits_{j=1}^re^{2i\pi\alpha'
(\alpha_i,\beta_j)}A_n(1,\{\sigma\},n)\right],\label{KK-string}\eea
where we have defined $\alpha_0=1$. Our notations are following.
$O\{\alpha\}$ means to keep the relative ordering inside the set
$\alpha$ while $\alpha^T$ means to take the reversed ordering of set
$\alpha$. Putting together
 $P(O\{\alpha\}\cup
O\{\beta^T\})$ denote all  permutations of set $O\{\alpha\}\cup
O\{\beta^T\}$ where relative orderings inside set $\alpha$ and set
$\beta^T$ have been preserved. This relation expresses $n=r+s+2$
point open string tree amplitudes by $(n-2)!$ amplitudes.

 BCJ relation for open string tree amplitudes is given
 as\cite{BjerrumBohr:2009rd}
 \bea \mathcal{I}m\left[\prod\limits_{1\leq
i<j\leq r}e^{2i\pi\alpha'k_{\beta_i}\cdot
k_{\beta_j}}\Sl_{\{\sigma\}\in P(O\{\alpha\}\cup
O\{\beta^T\})}\prod\limits_{i=0}^s\prod\limits_{j=1}^re^{2i\pi\alpha'(\alpha_i,\beta_j)}
A_n(1,\{\sigma\},n)\right]=0,\label{BCJ-string}
 \eea
 where $(\alpha,\beta)$ is defined as
 \bea
 (\alpha,\beta)=\Biggl\{
                  \begin{array}{cc}
                    k_{\alpha}\cdot k_{\beta} & (x_{\beta}>x_{\alpha}) \\
                    0 & \text{otherwise} \\
                  \end{array}.
 \eea
 Combining with KK-relation (\ref{KK-string}), this relation reduces further the number
  of
 independent open string amplitudes from $(n-2)!$ to $(n-3)!$.

Noticing that
  \bea & &\prod\limits_{1\leq i<j\leq
r}e^{2i\pi\alpha'k_{\beta_i}\cdot k_{\beta_j}}\Sl_{\{\sigma\}\in
P(O\{\alpha\}\cup
O\{\beta^T\})}\prod\limits_{i=0}^s\prod\limits_{j=1}^re^{2i\pi\alpha'(\alpha_i,\beta_j)}A_n(1,\{\sigma\},n)\nn
&=&\Sl_{\{\sigma\}\in P(O\{\alpha\}\cup O\{\beta^T\})}\exp
\left(i\pi\alpha'\Sl_{1\leq i<j\leq
r}s_{\beta_i\beta_j}+\Sl_{i=0}^s\Sl_{j=1}^ri\pi\alpha'(\alpha_i,\beta_j)\right)A_n(1,\{\sigma\},n)\nn
&=&\Sl_{\{\sigma\}\in P(O\{\alpha\}\cup O\{\beta^T\})}\exp
\left(i\pi\alpha'\Sl_{i=1}^r\Sl_{\sigma_J<\sigma_{\beta_i}}s_{\beta_iJ}\right)A_n(1,\{\sigma\},n),
 \eea
 where $\sigma_J$ denotes the position of the open string $J$ in the permutation $\{\sigma\}$
 with the convention that the position of particle $1$ is
defined as $\sigma_1=0$, we can take the field theory limit
 $\alpha'\rightarrow 0$ where only massless modes (i.e., gluons) of  open
string theory are left. The leading contribution of the real part of Eq.
\eqref{KK-string} gives the familiar KK relation for color-ordered
gluon amplitudes in field theory
\bea
A_n(\beta_1,...,\beta_r,1,\alpha_1,...,\alpha_s,n)=(-1)^r\Sl_{\{\sigma\}\in
P(O\{\alpha\}\cup O\{\beta^T\})}A_n(1,\{\sigma\},n).
 \eea
The leading contribution  of the imaginary part of Eq.
\eqref{BCJ-string} gives
 \bea \Sl_{\{\sigma\}\in P(O\{\alpha\}\cup
O\{\beta\})}\Sl_{i=1}^r\Sl_{\sigma_J<\sigma_{\beta_i}}s_{\beta_iJ}A_n(1,\{\sigma\},n)=0,\label{gen-BCJ}
 \eea
where we have redefined $\{\beta\}^T\rightarrow \{\beta\}$ with
$r$-elements. This is nothing, but the BCJ relation for
color-ordered gluon amplitudes in field theory. When there is only
one element in $\{\beta\}$, it just gives the fundamental BCJ
relation, which has been proved in field theory \cite{Feng:2010my,
Tye:2010kg, Jia:2010nz}. The formula (\ref{gen-BCJ}) is more general
in following sense. We can divide remaining $(n-2)$-elements into
arbitrary two sets $\alpha,\beta$ with the number of elements
$(n-2-r),r$ respectively and with arbitrary ordering. Because this
freedom, remaining $(n-2)$-elements are all same footing.

As we have mentioned,  the BCJ relation\eqref{gen-BCJ} provides
further constraints on  amplitudes to reduce the number of
independent amplitudes from $(n-2)!$ to $(n-3)!$. However,
\eqref{gen-BCJ}  provides only a set of constraints on amplitudes,
and it does not give the explicit expressions of amplitudes by
minimal-basis. To obtain these explicit expressions, one should
solve these constraint equations to express any amplitudes by
$(n-3)!$ independent ones. However, the general solving is very
nontrivial and based on some examples, an explicit minimal-basis
expression  was conjectured in \cite{Bern:2008qj}. In  following two
sections, we will first give the field theory proof of the
constraint equations \eqref{gen-BCJ},
 then we will use these constraints to prove (solve)
 the conjectured explicit minimal-basis expression.

\section{The field theory proof of the general  BCJ relation}\label{second-proof}

Though the general  BCJ relation \eqref{gen-BCJ} can be derived from
the BCJ relation in string theory directly by taking the field
theory limit, its pure field theory proof  is still desirable. In
this section, we will give a field theory proof of  the general BCJ
relation \eqref{gen-BCJ} by BCFW recursion relation. To demonstrate
our idea of proof, we show an example first.

\subsection{An example}

The first nontrivial example for the general BCJ relation is the six
point amplitude with set $\beta=\{2,3\}$ and set $\alpha=\{4,5\}$
and it is given by
\bea
0=I_6&=&(s_{21}+s_{31}+s_{32})A(1,2,3,4,5,6)+(s_{21}+s_{31}+s_{32}+s_{34})A(1,2,4,3,5,6)\nn
&+&(s_{21}+s_{31}+s_{32}+s_{34}+s_{35})A(1,2,4,5,3,6)+(s_{21}+s_{24}+s_{31}+s_{34}+s_{32})A(1,4,2,3,5,6)\nn
&+&(s_{21}+s_{24}+s_{31}+s_{34}+s_{32}+s_{35})A(1,4,2,5,3,6)\nn
&+&(s_{21}+s_{24}+s_{25}+s_{31}+s_{34}+s_{35}+s_{32})A(1,4,5,2,3,6)\label{6pt-BCJ}
 \eea
To show this is true, there are two possible ways to go. The first
way is to use the fundamental BCJ relation
recursively\cite{Feng:2010my}, while the second way, the BCFW
recursion relation\cite{Britto:2004ap,Britto:2005fq}. In this paper
we will use the second method where
 $p_1$ and $p_6$ are the shifted momenta. Thus
the R. H. S. of the above equation  is
 \bea
&&(s_{21}+s_{31}+s_{32})\left[A(\WH 1,2|3,4,5,\WH 6)+A(\WH
1,2,3|4,5,\WH 6)+A(\WH 1,2,3,4|5,\WH 6)\right]\nn
&+&(s_{21}+s_{31}+s_{32}+s_{34})\left[A(\WH 1,2|4,3,5,\WH 6)+A(\WH
1,2,4|3,5,\WH 6)+A(\WH 1,2,4,3|5,\WH 6)\right]\nn
&+&(s_{21}+s_{31}+s_{32}+s_{34}+s_{35})\left[A(\WH 1,2|4,5,3,\WH
6)+A(\WH 1,2,4|5,3,\WH 6)+A(\WH 1,2,4,5|3,\WH 6)\right]\nn
&+&(s_{21}+s_{24}+s_{31}+s_{34}+s_{32})\left[A(\WH 1,4|2,3,5,\WH
6)+A(\WH 1,4,2|3,5,\WH 6)+A(\WH 1,4,2,3|5,\WH 6)\right]\nn
&+&(s_{21}+s_{24}+s_{31}+s_{34}+s_{32}+s_{35})\left[A(\WH
1,4|2,5,3,\WH 6)+A(\WH 1,4,2|5,3,\WH 6)+A(\WH 1,4,2,5|3,\WH
6)\right]\nn
&+&(s_{21}+s_{24}+s_{25}+s_{31}+s_{34}+s_{35}+s_{32})\left[A(\WH
1,4|5,2,3,\WH 6)+A(\WH 1,4,5|2,3,\WH 6)+A(\WH 1,4,5,2|3,\WH
6)\right].
 \eea
where we have used following convention $A(\WH 1,4,5,2|3,\WH
6)\equiv { \sum_h A_L(\WH 1,4,5,2,\WH P_{3,\WH 6}^{h}) A_R(-\WH
P_{3,\WH 6}^{-h},3,\WH 6)\over s_{36}}$. Since the momentum of $k_1$
in the BCFW expansion of amplitudes is shifted, when we use the BCJ
relation for sub-amplitudes $A_L$ or $A_R$, we should use the
kinematic factors $s_{2\WH 1}$ and $s_{3\WH 1}$ instead of $s_{21}$
and $s_{31}$. Thus we should write  $s_{21}= s_{2\WH 1}+ (s_{21}-
s_{2\WH 1})$ etc. Putting it back, above expression  can be split
into two parts $\mathbb{A}$ and $\mathbb{B}$.
 $\mathbb{A}$ part contains  terms with kinematic factors $s_{2\WH1}$, $s_{3\WH1}$
 and $s_{ij}$($i,j\neq 1$), while $\mathbb{B}$ part
 contains  terms with  kinematic factors $s_{21}-s_{2\WH 1}$ and $s_{31}-s_{3\WH 1}$.

The $\mathbb{A}$ part can  be rewritten as
 \bea&&\mathbb{A}=\Bigl[\left(s_{2\WH 1}+s_{3\WH P_{12}}\right)A(\WH 1,2|3,4,5,\WH 6)+\left(s_{2\WH 1}+s_{3\WH P_{12}}+s_{34}\right)A(\WH 1,2|4,3,5,\WH
 6)\nn
 &&+\left(s_{2\WH 1}+s_{3\WH P_{12}}+s_{34}+s_{35}\right)A(\WH 1,2|4,5,3,\WH
6)\Bigr]+\left[\left(s_{2\WH 1}+s_{3\WH 1}+s_{32}\right)A(\WH
1,2,3|4,5,\WH 6)\right]\nn &&+\left[\left(s_{2\WH 1}+s_{3\WH
1}+s_{32}\right)A(\WH 1,2,3,4|5,\WH 6)\right]+\Bigl[\left(s_{2\WH
P_{14}}+s_{3\WH P_{14}}+s_{32}\right)A(\WH 1,4|2,3,5,\WH 6)\nn
&&+\left(s_{2\WH P_{14}}+s_{3\WH P_{14}}+s_{32}+s_{35}\right)A(\WH
1,4|2,5,3,\WH 6)+\left(s_{2\WH P_{14}}+s_{25}+s_{3\WH
P_{14}}+s_{35}+s_{32}\right)A(\WH 1,4|5,2,3,\WH 6)\Bigr]\nn
&&+\Bigl[\left(s_{2\WH 1}+s_{3\WH P_{124}}\right)A(\WH 1,2,4|3,5,\WH
6)+\left(s_{2\WH 1}+s_{24}+s_{3\WH P_{142}}\right)A(\WH
1,4,2|3,5,\WH 6)\nn &&+\left(s_{2\WH 1}+s_{3\WH
P_{124}}+s_{35}\right)A(\WH 1,2,4|5,3,\WH 6)+\left(s_{2\WH
1}+s_{24}+s_{3\WH P_{142}}+s_{35}\right)A(\WH 1,4,2|5,3,\WH
6)\Bigr]\nn &&+\Bigl[\left(s_{2\WH P_{145}}+s_{3\WH
P_{145}}+s_{32}\right)A(\WH 1,4,5|2,3,\WH 6)\Bigr]\nn
&&+\Bigl[\left(s_{2\WH 1}+s_{3\WH 1}+s_{32}+s_{34}\right)A(\WH
1,2,4,3|5,\WH 6)+\left(s_{2\WH 1}+s_{24}+s_{3\WH
1}+s_{34}+s_{32}\right)A(\WH 1,4,2,3|5,\WH 6)\Bigr]\nn
&&+\Bigl[\left(s_{2\WH 1}+s_{3\WH P_{1245}}\right)A(\WH
1,2,4,5|3,\WH 6)+\left(s_{2\WH 1}+s_{24}+s_{3\WH
P_{1425}}\right)A(\WH 1,4,2,5|3,\WH 6)\nn &+&\left(s_{2\WH
1}+s_{24}+s_{25}+s_{3\WH P_{1452}}\right)A(\WH 1,4,5,2|3,\WH
6)\Bigr] \eea
For a given BCFW splitting (i.e., the particular cut, for example
$(12|3456)$) we should use the BCJ relation for $A_L$ or $A_R$
sub-amplitudes. For example, the splitting $(12)$ can be grouped as
  \bea
  &&\left(s_{2\WH 1}+s_{3\WH P_{12}}\right)A(\WH 1,2|3,4,5,\WH 6)+\left(s_{2\WH 1}+s_{3\WH P_{12}}+s_{34}\right)A(\WH 1,2|4,3,5,\WH
 6)\nn
 &&+\left(s_{2\WH 1}+s_{3\WH P_{12}}+s_{34}+s_{35}\right)A(\WH 1,2|4,5,3,\WH
6)\nn
&=&s_{2\WH 1}\left(A(\WH 1,2|3,4,5,\WH 6)+A(\WH 1,2|4,3,5,\WH
 6)+A(\WH 1,2|4,5,3,\WH
6)\right)\nn
&&+s_{3\WH P_{12}}A(\WH 1,2|3,4,5,\WH 6)+\left(s_{3\WH P_{12}}+s_{34}\right)A(\WH 1,2|4,3,5,\WH
 6)+\left(s_{3\WH P_{12}}+s_{34}+s_{35}\right)A(\WH 1,2|4,5,3,\WH
6)\nn
&=&0,
  \eea
 where we have used the  BCJ relation for three point amplitudes and five point amplitudes.
 Similar arguments can be used to show that
whole  $\mathbb{A}$ part vanishes.

The $\mathbb{B}$ part can be rewritten as
 \bea
&&\mathbb{B}=s_{21}\Big[A(1,2,3,4,5,6)+A(1,2,4,3,5,6)+A(1,2,4,5,3,6)
+A(1,4,2,3,5,6)\nn &&+A(1,4,2,5,3,6)+A(1,4,5,2,3,6)\Big]\nn
&&+s_{31}\Big[A(1,2,3,4,5,6)+A(1,2,4,3,5,6)+A(1,2,4,5,3,6)
+A(1,4,2,3,5,6)\nn &&+A(1,4,2,5,3,6)+A(1,4,5,2,3,6)\Big]\nn
&&+\oint_{z\neq 0}\frac{dz}{z}s_{2\WH 1}\Big[A(\WH 1,2,3,4,5,\WH
6)+A(\WH 1,2,4,3,5,\WH 6)+A(\WH 1,2,4,5,3,\WH 6) +A(\WH
1,4,2,3,5,\WH 6)\nn &&+A(\WH 1,4,2,5,3,\WH 6)+A(\WH 1,4,5,2,3,\WH
6)\Big]\nn && +\oint_{z\neq 0}\frac{dz}{z}s_{3\WH 1}\Big[A(\WH
1,2,3,4,5,\WH 6)+A(\WH 1,2,4,3,5,\WH 6)+A(\WH 1,2,4,5,3,\WH 6)
+A(\WH 1,4,2,3,5,\WH 6)\nn &&+A(\WH 1,4,2,5,3,\WH 6)+A(\WH
1,4,5,2,3,\WH 6)\Big]\nn
 &&=-\oint_{z=\infty}\frac{dz}{z}s_{2\WH
1}\Big[A(\WH 1,2,3,4,5,\WH 6)+A(\WH 1,2,4,3,5,\WH 6)+A(\WH
1,2,4,5,3,\WH 6) +A(\WH 1,4,2,3,5,\WH 6)\nn &&+A(\WH 1,4,2,5,3,\WH
6)+A(\WH 1,4,5,2,3,\WH 6)\Big]\nn
&&-\oint_{z=\infty}\frac{dz}{z}s_{3\WH 1}\Big[A(\WH 1,2,3,4,5,\WH
6)+A(\WH 1,2,4,3,5,\WH 6)+A(\WH 1,2,4,5,3,\WH 6) +A(\WH
1,4,2,3,5,\WH 6)\nn &&+A(\WH 1,4,2,5,3,\WH 6)+A(\WH 1,4,5,2,3,\WH
6)\Big],
  \eea
  where $\oint_{z\neq 0}$ means the big enough contour around $z=0$
  but does not include the contribution from pole $z=0$.
Using the KK relation, we can see
\bea &&A(\WH 1,2,3,4,5,\WH
6)+A(\WH 1,2,4,3,5,\WH 6)+A(\WH 1,2,4,5,3,\WH 6) +A(\WH
1,4,2,3,5,\WH 6)\nn &&+A(\WH 1,4,2,5,3,\WH 6)+A(\WH 1,4,5,2,3,\WH
6)\nn &&= A( \{3,2\},\WH 1,\{4,5\},\WH 6).\eea
Since in $A( \{3,2\},\WH 1,\{4,5\},\WH 6)$, the $\WH 1$ and $\WH 6$
are not nearby, the $z\rightarrow \infty$
behavior\cite{ArkaniHamed:2008yf} is $\frac{1}{z^2}$, the integrals
in $\mathbb{B}$ must vanish. Thus finally we proved the BCJ relation
for six gluon amplitudes \eqref{6pt-BCJ}. It is worth to notice that
the method of proof is similar to the one used for the fundamental
BCJ relation\cite{Feng:2010my}.

In the next subsection, we will extend the proof of this example to
the  proof of $n$-particle  BCJ relation \eqref{gen-BCJ} with
arbitrary sets $\alpha,\beta$.

\subsection{General proof}

Now let us turn to the field theory proof of the general  BCJ
relation \eqref{gen-BCJ}. The starting point of the recursive proof
is the  BCJ relation for three point amplitudes
\bea
s_{21}A_3(1,2,3)=0. \eea
This relation can be seen obviously, since $s_{21}=p_3^2=0$.

For the general formula where the momenta of the legs $1$ and $n$
are the shifted momenta in the BCFW expansion, we should notice that
there are two types of dynamical factors $s_{ij}$: one contains the
shifted momentum $p_1$ and another, does not. As seen in previous
example, these two types should be treated separately when we use
BCFW recursion relation to expand amplitudes and when we  apply
general BCJ relation inductively to sub-amplitudes,
 where we should use the shifted factors $s_{\beta_i\WH 1}$, i.e., we should write
 $s_{\beta_i1}=s_{\beta_i\WH 1}+\left(s_{\beta_i1}-s_{\beta_i\WH 1}\right)$.

 As in the previous example, we can divide the expression into two parts: part $\mathbb{A}$
 and the part $\mathbb{B}$. The contribution of part $\mathbb{B}$ with
 factors
 $\left(s_{\beta_i1}-s_{\beta_i\WH 1}\right)$
 is given as
  \bea & &\Sl_{\{\sigma\}\in
P(O\{\alpha\}\cup O\{\beta\})}\Sl_{i=1}^rs_{\beta_i1}A( 1,\{\sigma\},
n)-\Sl_{\{\sigma\}\in P(O\{\alpha\}\cup
O\{\beta\})}\Sl_{i=1}^r\Sl_{\text{All splitting}}s_{\beta_i\WH 1}A(\WH 1,\{\sigma_L\}|\{\sigma_R\},\WH n)\nn
&=&\Sl_{i=1}^rs_{\beta_i1}\Sl_{\{\sigma\}\in P(O\{\alpha\}\cup
O\{\beta\})}A( 1,\{\sigma\}, n)+\Sl_{i=1}^r\Sl_{\{\sigma\}\in
P(O\{\alpha\}\cup O\{\beta\})}\oint_{big~z\neq
0}\frac{dz}{z}s_{\beta_i\WH 1}A(\WH 1,\{\sigma\},\WH n)\nn
&=&\Sl_{i=1}^r\Sl_{\{\sigma\}\in P(O\{\alpha\}\cup
O\{\beta\})}\left[-\oint_{z=\infty}\frac{dz}{z}s_{\beta_i\WH 1}A(\WH
1,\{\sigma\},\WH n)\right]\nn
&=&(-1)^{r+1}\Sl_{i=1}^r\oint_{z=\infty}\frac{dz}{z}s_{\beta_i\WH
1}A(\{\beta^T\},\WH 1,\{\alpha\},\WH n) ,
 \eea
where we have used the KK relation in the last line. Since $\WH 1$
and $\WH n$ are not nearby, the boundary behavior is
$\frac{1}{z^2}$\cite{ArkaniHamed:2008yf}, thus the integral around
the infinity  vanishes.

The contribution of part $\mathbb{A}$ of \eqref{gen-BCJ} is given by
following  BCFW recursion relation as
 \bea
&& \Sl_{\{\sigma\}\in P(O\{\alpha\}\cup
O\{\beta\})}\Sl_{i=1}^r\Sl_{\sigma_J<\sigma_{\beta_i}}s_{\beta_iJ}\Sl_{\text{All
splitting}}A_n(\WH 1,\{\sigma_L\}|\{\sigma_R\},\WH
n)\nn
&=&\Sl_{i=1}^r\Sl_{\sigma_J<\sigma_{\beta_i}}s_{\beta_iJ}\Sl_{\text{All
splitting}}\Sl_{\{\sigma_L\}\subset
P(O\{\alpha_L\}\cup O\{\beta_L\})}\Sl_{\{\sigma_R\}\in
P(O\{\alpha_R\}\cup O\{\beta_R\})}A_n(\WH 1,\{\sigma_L\}|\{\sigma_R\},\WH
n)\nn &=&\Sl_{\text{All splitting}}\Biggl\{\left[\Sl_{\sigma_L\in
P(O\{\alpha_L\}\cup O\{\beta_L\})}\Sl_{i=1}^{r_L}\Sl_{0\leq\sigma_J<\beta_{Li}}s_{\beta_{Li}J_L}A_{r_L+s_{L}+2}(\WH
1,\{\sigma_L\},-\WH P_{\mathcal {I}})\right]\times\frac{1}{P_{\mathcal
{I}}^2}\nn &\times&\left[\Sl_{\{\sigma_R\}\in
P(O\{\alpha_R\}\cup O\{\beta_R\})}A_{n-r_L-s_L-2}A(\WH P_{\mathcal
{I}},\{\sigma_R\},\WH n)\right]\nn &+& \left[\Sl_{\{\sigma_L\}\in
P(O\{\alpha_L\}\cup O\{\beta_L\})}A_{r_L+s_{L}+2}(\WH 1,\{\sigma_L\},-\WH
P_{\mathcal {I}})\right]\times\frac{1}{P_{\mathcal {I}}^2}\nn
&\times&\left[\Sl_{\{\sigma_R\}\in
P(O\{\alpha_R\}\cup O\{\beta_R\})}\Sl_{i=r_L+1}^r\Sl_{r_L+s_L<\sigma_J<\beta_{Ri}}\left(s_{\beta_{Ri},J_R}+s_{\beta_{R_i}(\WH
P_{\mathcal {I}})}\right)A_{n-r_L-s_L-2}A(\WH P_{\mathcal {I}},\{\sigma_R\},\WH
n)\right]\Biggr\},
 \eea
 where the kinematic factors with $J=1$ are shifted, i.e., $s_{\beta_i\WH 1}$.
 In the equation above, we have used $\{\alpha_L\}$, $\{\beta_L\}$ to denote the subsets
  of $\alpha$, $\beta$ at the left hand side and
  $\{\alpha_R\}$, $\{\beta_R\}$  the subsets  of $\alpha$, $\beta$ at  the right hand side
   respectively. $r_L$, $s_L$, $r_R$ and
  $s_R$ are  the numbers of elements in each subset
  $\{\beta_L\}$, $\{\alpha_L\}$, $\{\beta_R\}$ and $\{\alpha_R\}$ respectively.
  $\WH P_{\mathcal {I}}$ is the sum of the momenta  at the left hand side of
   a given splitting $\mathcal {I}$.
With the general BCJ relation for sub-amplitudes, each term at the
last equation is zero, thus we have shown the contribution of part
$\mathbb{A}$ is zero. Combining results from part $\mathbb{A}$ and
part $\mathbb{B}$  we proved general BCJ relation\eqref{gen-BCJ}.

 \section{The
proof of the explicit minimal-basis expansion of gauge field tree
amplitudes}\label{first-proof}

Although the general BCJ relation \eqref{gen-BCJ} provides a set of
constraint equations, which reduces the number of minimal basis from
$(n-2)!$ to $(n-3)!$, the explicit expression of other amplitudes by
the minimal basis is not manifest. A manifest expression is
 conjectured in \cite{Bern:2008qj}. In this section, we will
prove the conjecture explicitly.

 The conjectured minimal-basis expansion can be written  as\cite{Bern:2008qj}
 \bea
  A_n(1,\beta_1,...,\beta_r,2,\alpha_1,...,\alpha_{n-r-3},n)=\Sl_{\{\xi\}\in
  P(\{\beta\}\cup O\{\alpha\})}A_n(1,2,\{\xi\},n)\prod\limits_{k=1}^r\frac{\mathcal
  {F}^{\{\beta\},\{\alpha\}}(2,\{\xi\},n|k)}{s_{1,\beta_1,...,\beta_k}},\label{orig-BCJ}
  \eea
where  $P(\{\beta\}\cup O\{\alpha\})$ corresponds to all
permutations of $\{\beta\}\cup\{\alpha\}$ that maintain the relative
order of the set $\{\alpha\}$. The function $\mathcal
  {F}^{\{\beta\},\{\alpha\}}(2,\{\xi\},n|k)$ is defined as
   \bea
  \mathcal {F}^{\{\beta\},\{\alpha\}}(2,\{\xi\},n|k)&=&\left\{
                          \begin{array}{cc}
                             \Sl_{\xi_J>\xi_{\beta_k}}\mathcal {G}(\beta_k,J)&\text{ if $\xi_{\beta_{k-1}}<\xi_{\beta_k}$}\\
                             -\Sl_{\xi_J<\xi_{\beta_k}}\mathcal {G}(\beta_k,J)&\text{ if $\xi_{\beta_{k-1}}>\xi_{\beta_k}$}\\
                          \end{array}
                        \right\}\nn
                        &+&\left\{
                             \begin{array}{cc}
                               s_{1,\beta_1,...,\beta_k}&\text{  if $\xi_{\beta_{k-1}}<\xi_{\beta_k}<\xi_{\beta_{k+1}}$} \\
                               -s_{1,\beta_1,...,\beta_k}&\text{   if $\xi_{\beta_{k-1}}>\xi_{\beta_k}>\xi_{\beta{k+1}}$} \\
                               0&\text{   else} \\
                             \end{array}
                           \right\},~~\label{F-def}
  \eea
where $\xi_J$ stands for the position of the leg $J$ in the
permutation of $\xi$ and we should include $\xi_0\equiv
\alpha_0\equiv 2$. We define $\xi_{\beta_0}\equiv\infty$ and
$\xi_{\beta_{r+1}}\equiv0$. The function $\mathcal {G}$ is defined
by
\bea \mathcal
{G}(\beta_k,\beta_j)=\left\{
                                    \begin{array}{cc}
                                      s_{\beta_k\beta_j} & \text{if $k<j$} \\
                                      0 & \text{else} \\
                                    \end{array}
                                  \right\},
 \eea
 \bea \mathcal {G}(\beta_k,\alpha_j)= s_{\beta_k\alpha_j},
 \eea
where  $\alpha_0=2$, $\alpha_{n-r-2}=n$, $\xi_2=\xi_{\alpha_0}\equiv0$, $\xi_{n}=\xi_{\alpha_{n-r-2}}\equiv n-2$.

 In this formula, the amplitude with
$\beta_1,\beta_2,...,\beta_r$ between $1$ and $2$ are expressed by
the basis amplitudes with no $\beta$ between $1$ and $2$.
Thus the formula \eqref{orig-BCJ} expresses  any amplitude by
$(n-3)!$ independent amplitudes with fixed positions of $1,2,n$
explicitly. We will discuss and prove the minimal-basis expansion
\eqref{orig-BCJ} in the following subsections.

\subsection{The properties of the function $\mathcal
  {F}$}\label{F-pro}

Before we prove the minimal-basis expression \eqref{orig-BCJ}, let us
 have a look at some useful properties of the function $\mathcal
  {F}$. It is worth to have some remarks from the definition (\ref{F-def}).
  First let us notice that the function has two groups of
  parameters. The first group of parameters is the up-index $\{\beta\},\{\alpha\}$,
  which provides the
  ordering information of amplitude for which we need to expand into basis.
  The second group of parameters is $(2,\{\xi\},n)$, which fixes the particular amplitude
  of the basis, and the number $k$ which tells us that it is the $k$-th kinematic
  factor coming from the $k$-th element $\beta_k$ of up-index. The
  second point we need to notice is that for the $k$-th kinematic
  factor, the key information we need is the permutation $\xi$ and
  the relative ordering between $\beta_k$ and $\beta_j, j\geq k-1$.
  If these information is same, we may get same $k$-th kinematic
  factor.

Having observed above points, let us consider the minimal-basis
expansion of following two amplitudes. The first one is
$A_n(1,\beta_1,...,\beta_r,2,\alpha_1,...,\alpha_{n-r-3},n)$
given by (\ref{orig-BCJ}). The second one is following expansion
  \bea
 && A_n(1,\beta_1,...,\beta_p,2,\gamma_1,...,\gamma_{n-p-3},n)\nn
  &=&\Sl_{\xi\in
  P(\{\beta_1,...,\beta_p\}\cup O\{\gamma_1,...,\gamma_{n-p-3}\})}A_n(1,2,\{\xi\},n)\prod\limits_{k=1}^p\frac{\mathcal
  {F}^{\{\beta_1,...,\beta_p\},\{\gamma_1,...,\gamma_{n-p-3}\}}(2,\{\xi\},n|k)}{s_{1,\beta_1,...,\beta_p}},
  \label{F-pro-1}
  \eea
  where $p\leq r$ and $\{\gamma_1,...,\gamma_{n-p-3}\}\in P (O\{\alpha\}\cup
  O\{\beta_{p+1},...,\beta_{r}\})$, which give the relation
  between these two would-be-expanded amplitudes.


  For $p<r$, from the definition of $\mathcal
  {F}$ (\ref{F-def}) we can see
    \bea {F}^{\{\beta_1,...,\beta_p\},\{\gamma_1,...,\gamma_{n-p-3}\}}(2,\{\xi\},n|k)
    =\mathcal{F}^{\{\beta_1,...,\beta_r\},\{\alpha_1,...,\alpha_{n-r-3}\}}(2,\{\xi\},n|k),~~~ (1\leq k<p),~~~
    \label{F-pro-2-1}\eea
which is obvious from the definition of (\ref{F-def}). The boundary
case $k=p$ is more complicated and is given by
  \bea
  \mathcal
  {F}^{\{\beta_1,...,\beta_p\},\{\gamma\}}(2,\{\xi\},n|p)=\Biggl\{
                                        \begin{array}{ll}
\mathcal{F}^{\{\beta_1,...,\beta_r\},\{\alpha_1,...,\alpha_{n-r-3}\}}(2,\{\xi\},n|p) &  \\
\mathcal{F}^{\{\beta_1,...,\beta_r\},\{\alpha_1,...,\alpha_{n-r-3}\}}(2,\{\xi\},n|p)-s_{1\beta_1,...,\beta_p},
& \xi_{p-1}<\xi_p<\xi_{p+1}
\\
\mathcal{F}^{\{\beta_1,...,\beta_r\},\{\alpha_1,...,\alpha_{n-r-3}\}}(2,\{\xi\},n|p)-s_{1\beta_1,...,\beta_p},
& \xi_{p-1}>\xi_p, \xi_{p+1}>\xi_p
                                        \end{array}\label{F-pro-2}
  \eea
according to different relative orderings among
$\xi_{p-1},\xi_p,\xi_{p+1}$.


Now we consider a given permutation $\xi$ in which the relative
order of beta is $\beta_l,\beta_{l+1},...,\beta_{r}$ for a given
$1\leq l<r$, i.e, we have
$\xi_{\beta_l}<\xi_{\beta_{l+1}}<...<\xi_{\beta_r}$. Furthermore we
assume  $\xi_{\beta_{l-1}}>\xi_{\beta_l}$. With these orderings,
following equations can be seen from the definition of $\mathcal
  {F}$ given by (\ref{F-def}):
  \bea
\mathcal{F}^{\{\beta_1,...,\beta_r\},\{\alpha\}}(2,\{\xi\},n|k)=\left\{
                                        \begin{array}{ll}
                                        \Sl_{\xi_J>\xi_{\beta_k}}\mathcal{G}(\beta_k,J)+s_{1,\beta_1,...,\beta_k}
=-\Sl_{\xi_J<\xi_{\beta_k}}\mathcal{G}(\beta_k,J)+s_{1,\beta_1,...,\beta_{k-1}}& (l<k<r) \\
                                          \Sl_{\xi_J>\xi_{\beta_k}}\mathcal{G}(\beta_k,J)
                                          =-\Sl_{\xi_J<\xi_{\beta_k}}\mathcal{G}(\beta_k,J)
                                          -s_{\beta_r1}-\sum_{j=1}^{r-1} s_{\beta_r \beta_j} & (k=r)  \\
                                          -\Sl_{{\xi_J<\xi_{\beta_k}}}\mathcal{G}(\beta_k,J) & (k=l) \\
                                        \end{array}\right.
                                        \label{F-pro-3}
 \eea
 where we have used the momentum conservation for the first and
 second lines and the rewriting $s_{1,\beta_1,...,\beta_k}=
 s_{1,\beta_1,...,\beta_{k-1}}+s_{1\beta_k}+\sum_{i=1}^{k-1}
 s_{\beta_i \beta_k}$.

The last case we want to discuss is the case with $\xi_{r-1}>\xi_r$.
Then by the definition (\ref{F-def}) we obtain
\bea
\mathcal{F}^{\{\beta_1,...,\beta_r\},\{\alpha\}}(2,\{\xi\},n|k=r)=
-\Sl_{{\xi_J<\xi_{\beta_r}}}\mathcal{G}(\beta_r,J)-s_{1,\beta_1,...,\beta_r}.\label{F-pro-4}
\eea

Properties (\ref{F-pro-2-1}),(\ref{F-pro-2}),(\ref{F-pro-3}) and
(\ref{F-pro-4}) are all we need for our proof. It can be briefly
seen as following. With the general  BCJ relation\eqref{gen-BCJ}, we
can express any amplitude with $\beta_1,...,\beta_r$ between $1$ and
$2$ by amplitudes with less $\beta$s between $1$ and $2$. Then we
use minimal-basis expansion to express these amplitudes with less
$\beta$s between $1$ and $2$ by those with all $\beta$s between $2$
and $n$. For an given amplitude belongs to the minimal basis, there
are several contributions to the coefficient, thus we need to
 use  properties of $\mathcal{F}$ to combine them together. By this way we can prove the
minimal basis expansion of amplitude with $r$ $\beta$s between $1$
and $2$ by induction. In the next subsection, we will use some
examples to demonstrate  this pattern.

\subsection{Examples}

In this subsection, we will give some examples to see the explicit minimal basis expansion of amplitudes.
The first and the simplest example is the minimal-basis expansion of amplitudes with only one element in $\{\beta\}$.
 The second one
is the case with two elements in $\{\beta\}$, the third one is the
case with three elements in $\{\beta\}$. Through these examples, the
idea of general proof will be more clear.

\subsubsection{Only one element in $\{\beta\}$}

If there is only one element in $\{\beta\}$, with the general
formula of BCJ relation \eqref{gen-BCJ} we can express the amplitude
with $\beta_1$ between $1$ and $2$ by those with $\beta_1$ between
$2$ and $n$ immediately as\footnote{Please remember  the convention
that $\xi_0=2=\alpha_0$ should be included in the sum over $\xi$.
This convention will be used for all calculations later.}
 \bea & & A_n(1,\beta_1,2,\alpha_1,...,\alpha_{n-4},n)\nn
&=&-\Sl_{\{\xi\}\in
P(O\{\beta_1\}\cup O\{\alpha\})}\frac{\left(s_{1\beta_1}+\Sl_{\xi_J<\xi_{\beta_1}}s_{\beta_1J}\right)}{s_{1\beta_1}}A(1,2,\{\xi\},n)\nn
&=&\Sl_{\{\xi\}\in
P(\{\beta_1\}\cup O\{\alpha\})}\frac{\mathcal{F}^{\{\beta_1\},\{\alpha\}}(2,\{\xi\},n|1)}{s_{1\beta_1}}A_n(1,2,\{\xi\},n),
 \eea
 where we have used the definition \eqref{F-def} with the
 relative ordering $\xi_{\beta_0}=\infty>\xi_{\beta_1}>\xi_{\beta_{r+1}}=0$.
 Since there is only one $\beta$,
the ordered set  of permutations $P(O\{\beta_1\}\cup O\{\alpha\})$ becomes
the partially ordered set of permutation $P(\{\beta_1\}\cup O\{\alpha\})$
 that maintain
the order of the $\{\alpha\}$ elements. It is the same result with
the minimal-basis expression of amplitudes with only one element in
$\{\beta\}$\cite{Bern:2008qj}. This example is nothing but the
fundamental BCJ relation.

\subsubsection{Two elements in $\{\beta\}$}

The amplitude with two elements in $\{\beta\}$ is the first
nontrivial example. If there are two elements in the set
$\{\beta\}$, from the general formula of BCJ relation
\eqref{gen-BCJ}, we can see that
\bea &&A_n(1,\beta_1,\beta_2,2,\alpha_1,...,\alpha_{n-5},n)\nn
&=&-\Sl_{\{\W\xi\}\in P(O\{\alpha\}\cup
O\{\beta_2\})}\frac{\left(s_{1\beta_1}+s_{1\beta_2}+s_{\beta_2\beta_1}+\Sl_{\W\xi_J<\W\xi_{\beta_2}}
s_{\beta_2J}\right)}{s_{1\beta_1\beta_2}}A_n(1,\beta_1,2,\{\W\xi\},n)\nn
&& -\Sl_{\{\xi\}\in P(O\{\alpha\}\cup
O\{\beta_1,\beta_2\})}\frac{\left(s_{1\beta_1}+s_{1\beta_2}+\Sl_{i=1}^2\Sl_{\xi_J<\xi_{\beta_i}}
s_{\beta_iJ}\right)}{s_{1\beta_1\beta_2}}A(1,2,\{\xi\},n)\nn
&=&-\Sl_{\{\W\xi\}\in P(O\{\alpha\}\cup
O\{\beta_2\})}\frac{\left(s_{1,\beta_1,\beta_2}+\Sl_{\W\xi_J<\W\xi_{\beta_2}}
\mathcal{G}(\beta_2,J)\right)}{s_{1\beta_1\beta_2}} \Sl_{\{\xi\}\in
P(\{\beta_1\}\cup
O\{\W\xi\})}\frac{\mathcal{F}^{\{\beta_1\},\{\W\xi\}}(2,\{\xi
 \},n|1)}{s_{1\beta_1}}A_n(1,2,\{\xi\},n)\nn &&-\Sl_{\{\xi\}\in
P(O\{\alpha\}\cup
O\{\beta_1,\beta_2\})}\frac{\left(s_{1,\beta_1,\beta_2}
+\Sl_{i=1}^2\Sl_{\xi_J<\xi_{\beta_i}}\mathcal{G}(\beta_i,J)\right)}{s_{1\beta_1\beta_2}}A(1,2,\{\xi\},n),
~~~\label{beta=2-1} \eea
where we have used the minimal-basis expansion for
$A_n(1,\{\beta_1\},2,\{\W\xi\},n)$. For the double sum in the first
line, it is easy to see that double sum $\Sl_{\{\W \xi\}\in
P(O\{\alpha\}\cup O\{\beta_2\})}\Sl_{\{\xi\}\in P(\{\beta_1\}\cup
O\{\W\xi\})}$ can be written as a single sum $\Sl_{\{\xi\}\in
P(O\{\alpha\}\cup \{\beta_1,\beta_2\})}$. This single sum can be
written into following two sums: (1) case $\mathbb{A}\equiv
\Sl_{\{\xi\}\in P(O\{\alpha\}\cup O\{\beta_1,\beta_2\})}$; (2) case
$\mathbb{B}\equiv \Sl_{\{\xi\}\in P(O\{\alpha\}\cup
O\{\beta_2,\beta_1\})}$. Case $\mathbb{A}$ needs to combine with the
second line of Eq. (\ref{beta=2-1}) while the case $\mathbb{B}$ is
independent itself.

Let us start from the case $\mathbb{B}$  with the ordering
$\xi_{\beta_1}>\xi_{\beta_2}$.  From the definition of
$\mathcal{G}$, we know that $\mathcal{G}(\beta_2,J)$ is independent
of the position of $\beta_1$, thus the factor
$-\left(s_{1,\beta_1,\beta_2}+\Sl_{\W\xi_J<\W\xi_{\beta_2}}
\mathcal{G}(\beta_2,J)\right)$ is nothing, but
$\mathcal{F}^{\{\beta\},\{\alpha\}}(2,\{\xi\},n|2)$. Similarly by
using the properties (\ref{F-pro-2}), the second factor
${F}^{\{\beta_1\},\{\W\xi\}}(2,\{\gamma\},n|1)={F}^{\{\beta_1,\beta_2\},\{\alpha\}}(2,\{\xi\},n|1)$.
Combining all together we see that the contribution of case
$\mathbb{B}$ is
\bea \Sl_{\{\xi\}\in P(O\{\alpha\}\cup
O\{\beta_2,\beta_1\})}\frac{\mathcal{F}^{\{\beta\},\{\alpha\}}(2,\{\xi\},n|2)}{s_{1\beta_1\beta_2}}
\frac{\mathcal{F}^{\{\beta\},\{\alpha\}}(2,\{\xi\},n|1)}{s_{1\beta_1}}A_n(1,2,\{\xi\},n).~~~\label{2beta-1-1}
 \eea

 For the ordering
$\xi_{\beta_1}<\xi_{\beta_2}$, the contribution  is given by
 \bea
&&-\Sl_{\{\xi\}\in P(O\{\alpha\}\cup O\{\beta_1,\beta_2\})}
\Biggl[\frac{\left(-\mathcal{F}^{\{\beta\},\{\alpha\}}(2,\{\xi\},n|2)+s_{1\beta_1}\right)}{s_{1\beta_1\beta_2}}
\frac{\left(\mathcal{F}^{\{\beta\},\{\alpha\}}(2,\{\xi\},n|1)-s_{1\beta_1}\right)}{s_{1\beta_1}}\nn
&&+\frac{\left(-\mathcal{F}^{\{\beta\},\{\alpha\}}(2,\{\xi\},n|2)-\mathcal{F}^{\{\beta\},\{\alpha\}}(2,\{\xi\},n|1)+s_{1\beta_1}\right)}{s_{1\beta_1\beta_2}}\Biggr]
A_n(1,2,\{\xi\},n),
 \eea
where we have used the properties  \eqref{F-pro-2}, \eqref{F-pro-3}
and the definition of $\mathcal{G}$ again. The coefficients of
amplitudes with less than two $\mathcal{F}$s cancel out. Only the
term with two $\mathcal{F}$s is left
\bea \Sl_{\{\xi\}\in
P(O\{\alpha\}\cup
O\{\beta_1,\beta_2\})}\frac{\mathcal{F}^{\{\beta\},\{\alpha\}}(2,\{\xi\},n|2)}{s_{1\beta_1\beta_2}}
\frac{\mathcal{F}^{\{\beta\},\{\alpha\}}(2,\{\xi\},n|1)}{s_{1\beta_1}}A_n(1,2,\{\xi\},n).~~~\label{2beta-1-2}
\eea

Combining results given in (\ref{2beta-1-1}) and (\ref{2beta-1-2}),
we  see that for both ordering of $\beta_1,\beta_2$,  coefficients
are just those given in the minimal-basis expansion
\eqref{orig-BCJ}, thus   we have given the proof of minimal-basis
expansion with two $\beta$s.

\subsubsection{Three elements in $\{\beta\}$}

The next  nontrivial example is the minimal-basis expansion with
three elements in $\{\beta\}$. From the general formula of BCJ
relation \eqref{gen-BCJ}, we can express the amplitudes with three
$\beta$s by amplitudes with $\beta$s less than three as
\bea
&&A_n(1,\beta_1,\beta_2,\beta_3,2,\alpha_1,...,\alpha_{n-6},n)\nn
&=&-\Sl_{\{\xi\}\in P(O\{\alpha\}\cup
O\{\beta_3\})}\frac{s_{1,\beta_1,\beta_2,\beta_3}+\Sl_{\xi_J<\xi_{\beta_3}}s_{\beta_3J}}{s_{1,\beta_1,\beta_2,\beta_3}}A_n(1,\beta_1,\beta_2,2,\{\xi\},n)\nn
&&-\Sl_{\{\xi\}\in P(O\{\alpha\}\cup
O\{\beta_2,\beta_3\})}\frac{s_{1\beta_1}+s_{1\beta_2}+s_{\beta_2\beta_1}+s_{1\beta_3}+s_{\beta_3\beta_1}+\Sl_{i=2}^3\Sl_{\xi_J<\xi_{\beta_i}}s_{\beta_iJ}}{s_{1,\beta_1,\beta_2,\beta_3}}A_n(1,\beta_1,2,\{\xi\},n)\nn
&&-\Sl_{\{\xi\}\in P(O\{\alpha\}\cup
O\{\beta_1,\beta_2,\beta_3\})}\frac{s_{1\beta_1}+s_{1\beta_2}+s_{1\beta_3}+\Sl_{i=1}^3\Sl_{\xi_J<\xi_{\beta_i}}s_{\beta_iJ}}{s_{1,\beta_1,\beta_2,\beta_3}}A_n(1,2,\{\xi\},n).
 \eea
With the definition of $\mathcal{G}$ and the minimal-basis expansion with
$\beta$s less than three, the amplitude becomes \bea
&&A_n(1,\beta_1,\beta_2,\beta_3,2,\alpha_1,...,\alpha_{n-6},n)\nn
&=&-\Sl_{\{\xi\}\in
P(O\{\alpha\}\cup O\{\beta_3\})}\frac{s_{1,\beta_1,\beta_2,\beta_3}
+\Sl_{\xi_J<\xi_{\beta_3}}\mathcal{G}(\beta_3,J)}{s_{1,\beta_1,\beta_2,\beta_3}}\nn
&&\times\Sl_{\{\gamma\}\in P(\{\beta_1,\beta_2\}\cup O\{\xi\})}\frac{\mathcal{F}^{\{\beta_1,\beta_2\},\{\xi\}}(2,\{\gamma\},n|2)}{s_{1,\beta_1,\beta_2}}\frac{\mathcal{F}^{\{\beta_1,\beta_2\},\{\xi\}}(2,\{\gamma\},n|1)}{s_{1\beta_1}}A_n(1,2,\{\gamma\},n)\nn
&&-\Sl_{\{\xi\}\in
OP(O\{\alpha\}\cup O\{\beta_2,\beta_3\})}\frac{s_{1,\beta_1,\beta_2,\beta_3}
+\Sl_{i=2}^3\Sl_{\xi_J<\xi_{\beta_i}}\mathcal{G}(\beta_i,J)}{s_{1,\beta_1,\beta_2,\beta_3}}
\Sl_{\{\gamma\}\in P(\{\beta_1\}\cup O\{\xi\})}\frac{\mathcal{F}^{\{\beta_1\},\{\xi\}}(2,\{\gamma\},n|1)}{s_{1,\beta_1}}A_n(1,2,\{\gamma\},n)\nn
&&-\Sl_{\{\xi\}\in
P(O\{\alpha\}\cup O\{\beta_1,\beta_2,\beta_3\})}\frac{s_{1,\beta_1,\beta_2,\beta_3}
+\Sl_{i=1}^3\Sl_{\xi_J<\xi_{\beta_i}}\mathcal{G}(\beta_i,J)}{s_{1,\beta_1,\beta_2,\beta_3}}A_n(1,2,\{\xi\},n).\label{3-pt-eg}
 \eea
It can be seen that the first term gives $3!=6$ orderings among
$\beta_i$'s, while the second term gives only $3$ orderings and the
third term, only one.

To consider all these orderings, we  classify them  into following
three categories. The first one is $\xi_{\beta_2}>\xi_{\beta_3}$.
The second case is $\xi_{\beta_2}<\xi_{\beta_3}$ but $
\xi_{\beta_1}>\xi_{\beta_2}$. The third one is
$\xi_{\beta_1}<\xi_{\beta_2}<\xi_{\beta_3}$. We will discuss these
three categories one by one.

{\bf Case $\xi_{\beta_2}>\xi_{\beta_3}$:} For this case,  the second
and the third terms in \eqref{3-pt-eg} give no contributions. With
the properties of $\mathcal{F}$ and the definition of $\mathcal{G}$,
the first term of \eqref{3-pt-eg} gives
 \bea \Sl_{\{\xi\}\in
P(\{\beta_1,\beta_2,\beta_3\}\cup
O\{\alpha\}|\xi_{\beta_2}>\xi_{\beta_3})}&&
\frac{\mathcal{F}^{\{\beta\},\{\alpha\}}(2,\{\xi\},n|3)}{s_{1,\beta_1,\beta_2,\beta_3}}
\frac{\mathcal{F}^{\{\beta\},\{\alpha\}}(2,\{\xi\},n|2)}{s_{1,\beta_1,\beta_2}}
\frac{\mathcal{F}^{\{\beta\},\{\alpha\}}(2,\{\xi\},n|1)}{s_{1\beta_1}}\nn
&&\times A_n(1,2,\{\xi\},n).
 \eea
This gives the part with $\xi_{\beta_2}>\xi_{\beta_3}$ in the
minimal-basis expansion.

{\bf Case $\xi_{\beta_2}<\xi_{\beta_3}$ and $
\xi_{\beta_1}>\xi_{\beta_2}$:} For this case,  the third term of
\eqref{3-pt-eg} has no contribution. The combination of first and
the second terms of \eqref{3-pt-eg} gives
\bea &&-\Sl_{\xi\in P(\{\beta_1\}\cup
O\{\alpha\}\cup O\{\beta_2,\beta_3\}|\xi_{\beta_1}>\xi_{\beta_2})}
\Biggl[\frac{\left(-\mathcal{F}^{\{\beta\},\{\alpha\}}(2,\{\xi\},n|3)+s_{1,\beta_1,\beta_2}\right)}{s_{1,\beta_1,\beta_2,\beta_3}}
\frac{\left(\mathcal{F}^{\{\beta\},\{\alpha\}}(2,\{\xi\},n|2)-s_{1,\beta_1,\beta_2}\right)}{s_{1,\beta_1,\beta_2}}\nn
&&+\frac{\left(-\mathcal{F}^{\{\beta\},\{\alpha\}}(2,\{\xi\},n|3)-\mathcal{F}^{\{\beta\},\{\alpha\}}(2,\{\xi\},n|2)+s_{1,\beta_1,\beta_2}\right)}{s_{1,\beta_1,\beta_2,\beta_3}}
\Biggr]\frac{\mathcal{F}^{\{\beta\},\{\alpha\}}(2,\{\xi\},n|1)}{s_{1\beta_1}}A_n(1,2,\{\xi\},n),\nn
\eea
where we have used several properties of $\mathcal{F}$ presented in
section \ref{F-pro} and the definition of $\mathcal{G}$. In the
above expression, all the terms with $\mathcal{F}$s less than three
cancel out and  only term with three $\mathcal{F}$s is left
\bea
&&\Sl_{\xi\in P(\{\beta_1\}\cup O\{\alpha\}\cup
O\{\beta_2,\beta_3\}|\xi_{\beta_1}>\xi_{\beta_2})}
\frac{\mathcal{F}^{\{\beta\},\{\alpha\}}(2,\{\xi\},n|3)}{s_{1,\beta_1,\beta_2,\beta_3}}\frac{\mathcal{F}^{\{\beta\},\{\alpha\}}(2,\{\xi\},n|2)}{s_{1,\beta_1,\beta_2}}\nn
&&\times\frac{\mathcal{F}^{\{\beta\},\{\alpha\}}(2,\{\xi\},n|1)}{s_{1,\beta_1}}\times
A_n(1,2,\{\xi\},n). \eea
Thus it gives the contribution of the
permutations with $\xi_{\beta_2}<\xi_{\beta_3}$ but $
\xi_{\beta_1}>\xi_{\beta_2}$ in the minimal-basis expansion with
three $\beta$s.

{\bf Case $\xi_{\beta_1}<\xi_{\beta_2}<\xi_{\beta_3}$:} For this
one, all three terms contribute and we need to sum up. Using the
properties of $\mathcal{F}$ and the definition of $\mathcal{G}$
again, we have
{\small\bea &&-\Sl_{\{\xi\}\in P(O\{\alpha\}\cup
O\{\beta_1,\beta_2,\beta_3\})}
\Biggl[\frac{\left(-\mathcal{F}^{\{\beta\},\{\alpha\}}(2,\{\xi\},n|3)+s_{1,\beta_1,\beta_2}\right)}{s_{1,\beta_1,\beta_2,\beta_3}}
\frac{\left(\mathcal{F}^{\{\beta\},\{\alpha\}}(2,\{\xi\},n|2)-s_{1,\beta_1,\beta_2}\right)}{s_{1,\beta_1,\beta_2}}
\frac{\mathcal{F}^{\{\beta\},\{\alpha\}}(2,\{\xi\},n|1)}{s_{1,\beta_1}}\nn
&&+\frac{\left(-\mathcal{F}^{\{\beta\},\{\alpha\}}(2,\{\xi\},n|3)-\mathcal{F}^{\{\beta\},\{\alpha\}}(2,\{\xi\},n|2)+s_{1,\beta_1,\beta_2}+s_{1,\beta_1}\right)}{s_{1,\beta_1,\beta_2,\beta_3}}
\frac{\left(\mathcal{F}^{\{\beta\},\{\alpha\}}(2,\{\xi\},n|1)-s_{1\beta_1}\right)}{s_{1\beta_1}}\nn
&&+\frac{\left(-\mathcal{F}^{\{\beta\},\{\alpha\}}(2,\{\xi\},n|3)-\mathcal{F}^{\{\beta\},\{\alpha\}}(2,\{\xi\},n|2)-\mathcal{F}^{\{\beta\},\{\alpha\}}(2,\{\xi\},n|1)
+s_{1,\beta_1,\beta_2}+s_{1,\beta_1}\right)}{s_{1,\beta_1,\beta_2,\beta_3}}\Biggr]A_n(1,2,\{\xi\},n).\nn
\eea }
All the terms with $\mathcal{F}s$ less than three cancel out and we
are left with only term having  three $\mathcal{F}$s
\bea
\Sl_{\{\xi\}\in P(O\{\alpha\}\cup O\{\beta_1,\beta_2,\beta_3\})}
\frac{\mathcal{F}^{\{\beta\},\{\alpha\}}(2,\{\xi\},n|3)}{s_{1,\beta_1,\beta_2,\beta_3}}\frac{\mathcal{F}^{\{\beta\},\{\alpha\}}(2,\{\xi\},n|2)}{s_{1,\beta_1,\beta_2}}
\frac{\mathcal{F}^{\{\beta\},\{\alpha\}}(2,\{\xi\},n|1)}{s_{1,\beta_1}}A_n(1,2,\{\xi\},n).
 \eea
Summing over all three cases, we get the minimal-basis expansion
with three $\beta$s.

\subsection{General proof}

Having these examples above, we have seen the procedure of our
general proof. First we use  the general formula of BCJ
relation\eqref{gen-BCJ} to express the amplitude with $r$ $\beta$'s
by other amplitudes with  $\beta$'s less than $r$. Then using the
induction, we can substitute these amplitudes with $\beta$'s less
than $r$ by their explicit minimal-basis expansion. Next we divide
orderings of $r$ $\beta$'s into several cases and using the properties
of $\mathcal{F}$ and the definition of $\mathcal{G}$ to combine
contributions from various terms. The key  of the proof is (1) to
show the cancelation of these terms with ${\cal F}$'s less than $r$;
(2) the remaining term with $r$ ${\cal F}$'s are exact the one given
by the conjecture.  After summing over all  permutations we get the
minimal-basis expansion\eqref{orig-BCJ}.

Having above discussions, let us start with following expression
obtained by using the general formula of BCJ relation
\eqref{gen-BCJ}
 \bea &
&A_n(1,\beta_1,...,\beta_r,2,\alpha_1,...,\alpha_{n-r-3},n)\nn
&=&-\frac{1}{s_{1,\beta_1,...,\beta_r}}\Biggl[\Sl_{\{\xi\}\in
P(O\{\alpha\}\cup
O\{\beta_r\})}\left(s_{1,\beta_1,...,\beta_{r}}+\Sl_{\xi_J<\xi_{\beta_r}}s_{\beta_rJ}\right)A_n(1,\beta_1,...,\beta_{r-1},2,\{\xi\},n)\nn
&+&\Sl_{\{\xi\}\in P(O\{\alpha\}\cup
O\{\beta_{r-1},\beta_r\})}\left(s_{1,\beta_1,...,\beta_{r-1}}+s_{\beta_r\beta_{r-2}}+...+s_{\beta_r\beta_{1}}+s_{\beta_r1}+\Sl_{i=r-1}^r\Sl_{\xi_J<\xi_{\beta_i}}s_{\beta_iJ}\right)\nn
&&\times A_n(1,\beta_1,...,\beta_{r-2},2,\{\xi\},n)\nn &+&... \nn
&+&\Sl_{\{\xi\}\in P(O\{\alpha\}\cup
O\{\beta\})}\left(s_{1\beta_r}+...+s_{1\beta_1}+\Sl_{i=1}^r\Sl_{\xi_J<\xi_{\beta_i}}
s_{\beta_iJ}\right)A_n(1,2,\{\xi\},n)\Biggr].\label{recursion}\eea
Since in the right hand of above equation, all amplitudes have the
set $\beta$ with number less than $r$,  we can substitute the
minimal-basis expansions of these amplitudes into \eqref{recursion}
by induction. Thus we have
\bea
&&A_n(1,\beta_1,...,\beta_r,2,\alpha_1,...,\alpha_{n-r-3},n)\nn
&=&-\frac{1}{s_{1,\beta_1,...,\beta_r}}\Biggl[\Sl_{\{\xi\}\in
P(O\{\alpha\}\cup O\{\beta_r\})}\left(s_{1,\beta_1,...,\beta_{r}}
+\Sl_{\xi_J<\sigma_{\beta_r}}\mathcal{G}(\beta_r,J)\right)\nn
&&\times\Sl_{\{\gamma\}\in P(\{\beta_1,...,\beta_{r-1}\}\cup
O\{\xi\})}\prod_{k=1}^{r-1}\frac{\mathcal
  {F}^{\{\beta_1,...,\beta_{r-1}\},\{\xi\}}(2,\{\gamma\},n|k)}{s_{1,\beta_1,...,\beta_k}}A_n(1,2,\{\gamma\},n)\nn
  &+&\Sl_{\{\xi\}\in
P(O\{\alpha\}\cup O\{\beta_{r-1},\beta_r\})}
\left(s_{1,\beta_1,...,\beta_{r}}+\Sl_{i=r-1}^r\Sl_{\xi_J<\xi_{\beta_i}}\mathcal{G}(\beta_i,J)\right)\nn
&&\times\Sl_{\{\gamma\}\in P(\{\beta_1,...,\beta_{r-2}\}\cup O\{\xi\})}
\prod_{k=1}^{r-2}\frac{\mathcal
  {F}^{\{\beta_1,...,\beta_{r-2}\},\{\xi\}}(2,\{\gamma\},n|k)}{s_{1,\beta_1,...,\beta_k}}A_n(1,2,\{\gamma\},n)\nn
&+&...\nn &+&\Sl_{\{\xi\}\in
P(O\{\alpha\}\cup O\{\beta_1,...,\beta_r\})}\left(s_{1,\beta_1,...,\beta_r}+\Sl_{i=1}^r\Sl_{\xi_J<\xi_{\beta_i}}\mathcal{G}(\beta_i,J)\right)
A_n(1,2,\{\xi\},n)
 \Biggr].\label{induction}
 \eea

As we have seen in previous examples, the first term has $r!$
ordering and the second one, $(r-1)!$ and so on, until the last one,
only one ordering among $r$ $\beta$'s. We divide all orderings to
$r$ cases. The first case is the ordering with $\xi_{r-1}>\xi_r$,
which has contribution from  first term only and is given by
 \bea \Sl_{\xi\in
P(\{\beta_1,...,\beta_{r-1},\beta_r\}\cup
O\{\alpha\}|\xi_{r-1}>\xi_r)}\prod_{k=1}^{r}\frac{\mathcal
  {F}^{\{\beta\},\{\alpha\}}(2,\{\xi\},n|k)}{s_{1,\beta_1,...,\beta_k}}A_n(1,2,\{\xi\},n),\eea
  where we have already used \eqref{F-pro-4}.

Other orderings can be characterized  with
$\xi_{\beta_l}<\xi_{\beta_{l+1}}<...<\xi_{\beta_r}$ but
  $\xi_{\beta_{l-1}}>\xi_{\beta_l}$, where the value of $l$
  can be any integer between $1$ and $r$. For given $l$, only  first $r-l+1$ terms
   in \eqref{induction} give nonzero contributions.
  With the properties of $\mathcal{F}$ (especially (\ref{F-pro-3}))
   and the definition of $\mathcal{G}$, the terms with this ordering can be expressed by
   $\mathcal{F}^{\{\beta\},\{\alpha\}}$ as following
   \bea
   &-&\Sl_{\{\xi\}\in
P(\{\beta_1,...,\beta_{l-1}\}\cup O\{\alpha\}\cup
O\{\beta_l,...,\beta_r\}|\xi_{\beta_{l-1}}>\xi_{\beta_l})}
\frac{1}{s_{1,\beta_1,...,\beta_r}s_{1,\beta_1,...,\beta_{r-1}}...s_{1,\beta_1}}\nn
&\times&\Biggl\{\left(-\mathcal
  {F}^{\{\beta\},\{\alpha\}}(2,\{\xi\},n|r)+s_{1,\beta_1,...,\beta_{r-1}}\right)\left(\mathcal
  {F}^{\{\beta\},\{\alpha\}}(2,\{\xi\},n|r-1)-s_{1,\beta_1,...,\beta_{r-1}}\right)\nn
  &&\times\mathcal
  {F}^{\{\beta\},\{\alpha\}}(2,\{\xi\},n|r-2)\times...\times\mathcal
  {F}^{\{\beta\},\{\alpha\}}(2,\{\xi\},n|1)\nn
&+&\Sl_{k=l+1}^{r-1}\Biggl[\Bigl(-\mathcal
  {F}^{\{\beta\},\{\alpha\}}(2,\{\xi\},n|r)-\mathcal
  {F}^{\{\beta\},\{\alpha\}}(2,\{\xi\},n|r-1)-...-\mathcal
  {F}^{\{\beta\},\{\alpha\}}(2,\{\xi\},n|k)\nn
  &&+s_{1,\beta_1,...,\beta_{r-1}}+s_{1,\beta_{1},...,\beta_{r-2}}+...+s_{1,\beta_1,...,\beta_{k-1}}\Bigr)
 \Bigl(\mathcal
  {F}^{\{\beta\},\{\alpha\}}(2,\{\xi\},n|k-1)-s_{1,\beta_1,...,\beta_{k-1}}\Bigr)\nn
  &&\times \mathcal
  {F}^{\{\beta\},\{\alpha\}}(2,\{\xi\},n|k-2)\times...\times\mathcal
  {F}^{\{\beta\},\{\alpha\}}(2,\{\xi\},n|1) s_{1,\beta_{1},...,\beta_{r-1}}...s_{1,\beta_{1},...,\beta_k}\Biggr]\nn
&+& \Bigl(-\mathcal
  {F}^{\{\beta\},\{\alpha\}}(2,\{\xi\},n|r)-\mathcal
  {F}^{\{\beta\},\{\alpha\}}(2,\{\xi\},n|r-1)-...-\mathcal
  {F}^{\{\beta\},\{\alpha\}}(2,\{\xi\},n|l)\nn
  &+&s_{1,\beta_1,...,\beta_{r-1}}+s_{1,\beta_{1},...,\beta_{r-2}}+...+s_{1,\beta_1,...,\beta_l}\Bigr)\nn
  &&\times \mathcal
  {F}^{\{\beta\},\{\alpha\}}(2,\{\xi\},n|l-1)\times...\times\mathcal
  {F}^{\{\beta\},\{\alpha\}}(2,\{\xi\},n|1)s_{1,\beta_{1},...,\beta_{r-1}}...s_{1,\beta_{1},...,\beta_l} \Biggr\}
   .\label{cancelation}  \eea
Eq (\ref{cancelation}) is complicated, so we will discuss its
structure inside the big curly braked in details:
\begin{itemize}

\item (1) First it is easy to see that there are $r-l+1$ terms and
each term is given by the multiplication of various dynamical factor
$s$ and ${\cal F}$.  For the $k$-th term, there are $r+1-k$ factors
containing ${\cal F}$ and $k-1$ kinematic factors given by
$s_{1...\beta}$.

\item (2) The structure of kinematic factors is following. For the
first term, it is $1$. The second term has factor
$s_{1,\beta_1,...,\beta_{r-1}}$ and the third term,
$s_{1,\beta_1,...,\beta_{r-1}}s_{1,\beta_1,...,\beta_{r-2}}$. So for
$k$-th term, it is $\prod_{i=1}^{k-1}
s_{1,\beta_1,...,\beta_{r-i}}$.

\item (3) The structure of ${\cal F}$ is more complicated. Let us
use ${\cal F}_{t}\equiv {F}^{\{\beta\},\{\alpha\}}(2,\{\xi\},n|t)$,
then the first term is  $(-{\cal F}_r+s)({\cal F}_{r-1}-s)
\prod_{i=1}^{r-2} {\cal F}_{r-1-i}$. Especially only first two
${\cal F}$ combining with proper factor $s$. For the second term, we
merge first two ${\cal F}$ factors, i.e., $(-{\cal F}_r+s)({\cal
F}_{r-1}-s)\to (-{\cal F}_{r}-{\cal F}_{r-1}+s+s)$ while changing
next pure ${\cal F}_{r-2}$ factor into $({\cal F}_{r-2}-s)$. This
procedure is used recursively to the third, fourth terms and so on.
The only exception is for the last term, we do the merge only, but
not the changing.

Having above pattern, we can see that if we consider the power of
${\cal F}$, the first term contains power $r,r-1,r-2$, while the
second term, power $r-1,r-2,r-3$.  The observation can be summerized
by following $(r-l+1)\times (r-l+1)$ matrix:
\bea \left[\begin{array}{cccccccccc} r & r-1 & r-2 &  & & & & & & \\
0 & r-1 & r-2 & r-3 & & & & & &   \\  & & r-2 & r-3 & r-4 & & & & & \\
\cdot & \cdot& \cdot&\cdot &\cdot & \cdot& \cdot& \cdot& \cdot& \\ &
& & & & & & l+1 & l & l-1 \\ & & & & & & & & l & l-1
\end{array}\right]~~~\label{str}.\eea

\item (4) From previous observation, we see that for the $(r-k+1)$-th term, the first
two nontrivial factors have following structure. The first factor is
\bea & & \Bigl(-\mathcal
  {F}^{\{\beta\},\{\alpha\}}(2,\{\xi\},n|r)-\mathcal
  {F}^{\{\beta\},\{\alpha\}}(2,\{\xi\},n|r-1)-...-\mathcal
  {F}^{\{\beta\},\{\alpha\}}(2,\{\xi\},n|k)\nn
  &&+s_{1,\beta_1,...,\beta_{r-1}}+s_{1,\beta_{1},...,\beta_{r-2}}+...+s_{1,\beta_1,...,\beta_{k-1}}\Bigr),
  ~~~\label{factor-1}\eea
while the second factor is
\bea
 \Bigl(\mathcal
  {F}^{\{\beta\},\{\alpha\}}(2,\{\xi\},n|k-1)-s_{1,\beta_1,...,\beta_{k-1}}\Bigr)~~\label{factor-2}
  \eea

\end{itemize}

Having above structure (\ref{str}), now we can see how to finish our
proof. The whole expression has only one term with  $n$ factors
${\cal F}$, which is given  in the first term. It is nothing, but
the result we want to prove. Other contributions with less power of
${\cal F}$ will cancel each other.

To see the cancelation of various powers, let us start with the
power $(r-1)$, for which we need to sum up contributions from the
first and second terms. Recalling the factor
$s_{1,\beta_1,...,\beta_{r-1}}$ of second term, the contribution of
power $r-1$ from the second term is given by (where the factor
${\cal F}_{r-2}... {\cal F}_1$ is neglected)
\bea (-{\cal F}_{r}-{\cal F}_{r-1})s_{1,\beta_1,...,\beta_{r-1}}
\eea
while the contribution from the first term,
\bea (-{\cal F}_r+ s_{1,\beta_1,...,\beta_{r-1}}) ({\cal
F}_{r-1}-s_{1,\beta_1,...,\beta_{r-1}}) & \to & ({\cal F}_{r}+{\cal
F}_{r-1})s_{1,\beta_1,...,\beta_{r-1}}\eea

Next we move to the case with power $r-2$, where the first, second
and third terms contribute(where the factor ${\cal F}_{r-3}... {\cal
F}_1$ is neglected):
\bea First:~~~& -s_{1,\beta_1,...,\beta_{r-1}}^2  {\cal F}_{r-2}\nn
Second: ~~~ & s_{1,\beta_1,...,\beta_{r-1}} {\cal F}_{r-2}(
s_{1,\beta_1,...,\beta_{r-1}}+s_{1,\beta_1,...,\beta_{r-2}})\nn &
+s_{1,\beta_1,...,\beta_{r-1}}({\cal F}_{r}+{\cal
F}_{r-1})s_{1,\beta_1,...,\beta_{r-2}}\nn
Third: ~~~ &
s_{1,\beta_1,...,\beta_{r-1}}s_{1,\beta_1,...,\beta_{r-2}}(-{\cal
F}_r-{\cal F}_{r-1}-{\cal F}_{r-2})~~\label{r-2} \eea
From above table, it is easy to see the cancelation.

The cases of power $(r-1)$ and $(r-2)$ give, in fact, the general
pattern of cancelations. The general form of power $(h+1)$ of ${\cal
F}$  can be written as
\bea
 f(g,h) {\cal F}_{g} \prod_{i=1}^h {\cal F}_i,~~~\label{gen-F}
\eea
 where $g\geq h+1$ and  $ f(g,h)$ is a function of $s_{ij}$. As seen
 from the (\ref{factor-1}), (\ref{factor-2}) and (\ref{r-2}), we
 should divide the discussion of $g$ to two cases $g=h+1$ and $g>(h+1)$
 and for each case, we should carefully deal with the boundary
 given by the first and last term of (\ref{cancelation}).

{\bf The case $g>h+1$:} For this case, there are only two terms in
\eqref{cancelation} giving contributions, i.e., the terms with
$k=h+1$ and $k=h+2$. The term with $k=h+1$ gives a factor
$-s_{1,\beta_1,...,\beta_{r-1}}...s_{1,\beta_1,...,\beta_{h+1}}$
while the term with $k=h+2$ gives a factor
$s_{1,\beta_1,...,\beta_{r-1}}...s_{1,\beta_1,...,\beta_{h+1}}$.
Thus the sum of these two give $f(g,h)=0$ for $g>h+1$.

The boundary cases for $g>h+1$ should also be discussed
independently. For the upper boundary is nothing, but the one given
in (\ref{r-2}).
For the lower boundary, we should consider the case $h=l-1$, $g>l$.
In this case, the last term in \eqref{cancelation} gives a factor
$-s_{1,\beta_1,...,\beta_{r-1}}...s_{1,\beta_{1},...,\beta_l}$ while
the term with $k=l+1$ gives a factor
$s_{1,\beta_1,...,\beta_{r-1}}...s_{1,\beta_{1},...,\beta_l}$. These
two factors  cancel with each other. Thus we get $f(g,l-1)=0$ for
$g>l$.

{\bf The case $g=h+1$:} There are three terms  in
\eqref{cancelation} give contributions. The term $k=g=h+1$ gives a
factor
$-s_{1,\beta_1,...,\beta_{r-1}}...s_{1,\beta_1,...,\beta_{h+1}}$.
The them $k=g+1=h+2$ gives a factor
$(s_{1,\beta_1,...,\beta_{r-1}}+s_{1,\beta_1,...,\beta_{r-2}}+...+s_{1,\beta_1,...,\beta_{h+2}}+s_{1,\beta_1,...,\beta_{h+1}})s_{1,\beta_1,...,\beta_{r-1}}...s_{1,\beta_1,...,\beta_{h+2}}$.
The term $k=g+2=h+3$ gives a factor
$-(s_{1,\beta_1,...,\beta_{r-1}}+s_{1,\beta_1,...,\beta_{r-2}}+...+s_{1,\beta_1,...,\beta_{h+2}})s_{1,\beta_1,...,\beta_{r-1}}...s_{1,\beta_1,...,\beta_{h+2}}$.
Thus all three factors cancel out with each other, we have
$f(h+1,h)=0$.

Now we move to the boundary case. The upper boundary has been given
above.
For the lower boundary, two  cases $g=h+1=l$ and $g=h+1=l-1$ should
be considered.  For case $g=h+1=l$, the term with $k=l+2$ gives a
factor
$-(s_{1,\beta_1,...,\beta_{r-1}}+s_{1,\beta_1,...,\beta_{r-2}}+...+
s_{1,\beta_1,...,\beta_{l+2}}+s_{1,\beta_1,...,\beta_{l+1}})
s_{1,\beta_1,...,\beta_{r-1}}...s_{1,\beta_1,...,\beta_{l+1}}$,
while the term with $k=l+1$ gives a factor
$(s_{1,\beta_1,...,\beta_{r-1}}+s_{1,\beta_1,...,\beta_{r-2}}+...+
s_{1,\beta_1,...,\beta_{l+1}}+s_{1,\beta_1,...,\beta_{l}})s_{1,
\beta_1,...,\beta_{r-1}}...s_{1,\beta_1,...,\beta_{l+1}}$
 and the last term in \eqref{cancelation} gives a factor
 $-s_{1,\beta_1,...,\beta_{r-1}}...$ $s_{1,\beta_{1},...,\beta_{l+1}}s_{1,\beta_{1},...,\beta_{l}}$.
 Summing three terms up, we have  $f(l,l-1)=0$.
 For case $g=h+1=l-1$, the term with $k=l+1$ gives a factor
 $-(s_{1,\beta_1,...,\beta_{r-1}}+s_{1,\beta_1,...,\beta_{r-2}}+...+
 s_{1,\beta_1,...,\beta_{l+1}}+s_{1,\beta_1,...,\beta_{l}})
 s_{1,\beta_1,...,\beta_{r-1}}...s_{1,\beta_1,...,\beta_{l}}$
 while the last term in \eqref{cancelation} gives a factor
 $(s_{1,\beta_1,...,\beta_{r-1}}+s_{1,\beta_1,...,\beta_{r-2}}+...+
 s_{1,\beta_1,...,\beta_{l+1}}+s_{1,\beta_1,...,\beta_{l}})$
 $s_{1,\beta_1,...,\beta_{r-1}}...s_{1,\beta_1,...,\beta_{l}}$.
 Summing them up we get  $f(l-1,l-2)=0$.

Up to now, it is clear that all  terms with $\mathcal{F}$s less than
$r$ cancel out. Only the term with $r$ $\mathcal{F}$s is left. Thus
we  obtain
\bea
 &\Sl_{\{\xi\}\in
P(\{\beta_1,...,\beta_{l-1}\}\cup O\{\alpha\}\cup
O\{\beta_l,...,\beta_r\}|\xi_{\beta_{l-1}}>\xi_{\beta_l})}&
\frac{\mathcal
  {F}^{\{\beta\},\{\alpha\}}(2,\{\xi\},n|r)}{s_{1,\beta_1,...,\beta_r}}
  \frac{\mathcal
  {F}^{\{\beta\},\{\alpha\}}(2,\{\xi\},n|r-1)}{s_{1,\beta_1,...,\beta_{r-1}}}
  ...\nn
  &&\times\frac{\mathcal
  {F}^{\{\beta\},\{\alpha\}}(2,\{\xi\},n|1)}{s_{1\beta_1}}\times A_n(1,2,\{\xi\},n).
\eea
After summing over  all permutations, we get the expression of
explicit minimal-basis expansion \eqref{orig-BCJ}.

 \section{Summary}\label{conclusion}

 In this paper, we have done following two things. First we
gave a field theory proof of the general  BCJ relation obtained by
taking the field theory limit of imaginary part of monodromy
relation in string theory. Using this, we
  proved the explicit minimal-basis expansion of gauge field tree amplitudes which
was conjectured in \cite{Bern:2008qj}.

  \subsection*{Acknowledgements}
Y. J. Du would like to thank Q. Ma, J. L. Li and Z. Zhang for
many helpful discussions. Since most of his calculations are
finished at KITPC, Y. J. Du would also like to thank KITPC for
hospitality. Y. X. Chen and Y. J. Du are supported in part by the
NSF of China Grant No. 10775116, No. 11075138, and 973-Program Grant
No. 2005CB724508. BF is supported by fund from Qiu-Shi, the
Fundamental Research Funds for the Central Universities with
contract number 2010QNA3015, as well as Chinese NSF funding under
contract No.10875104.


\end{document}